\documentclass[conference,a4paper]{IEEEtran}
\usepackage{xcolor}
\usepackage{balance}
\usepackage{hyperref}
\usepackage{siunitx}
\usepackage{microtype}
\newcommand*{\colorJP}{\color{black}} % for easier way to control highlights
\ifCLASSINFOpdf
  \usepackage[pdftex]{graphicx}
\else
  \usepackage[dvips]{graphicx}
\fi
\usepackage{amsmath}
\ifCLASSOPTIONcompsoc
 \usepackage[caption=false,font=normalsize,labelfont=sf,textfont=sf]{subfig}
\else
 \usepackage[caption=false,font=footnotesize]{subfig}
\fi
\begin{document}
\title{DECT-2020 NR Link Distance Performance in Varying Environments: Models and Measurements}

\author{\IEEEauthorblockN{
Md Mohaiminul Haque\IEEEauthorrefmark{1},   % 1st author, 1st affiliations
Joonas Säe\IEEEauthorrefmark{1},   % 2nd author, 2nd affiliations
Juho Pirskanen\IEEEauthorrefmark{2},    % 3rd author, 3rd affiliations
and Mikko Valkama\IEEEauthorrefmark{1}      % 4th author, 4th affiliations
}                                     % ...
%\\
\IEEEauthorblockA{\IEEEauthorrefmark{1}Tampere Wireless Research Center, Tampere University, Finland, {mohaiminul.haque@tuni.fi}}
%\IEEEauthorblockA{\IEEEauthorrefmark{2}% 2nd affiliations
%Department of Information Technology and Communication Sciences, Tampere University, Finland, {joonas.sae@tuni.fi}}
\IEEEauthorblockA{\IEEEauthorrefmark{2}% 3rd affiliations
Wirepas Oy, Tampere, Finland, juho.pirskanen@wirepas.com}
%\IEEEauthorblockA{\IEEEauthorrefmark{4}% 4th affiliations
%Department of Information Technology and Communication Sciences, Tampere University, Finland, {jukka.lempiainen@tuni.fi}}  
% \IEEEauthorblockA{ \emph{*at least one e-mail address should be indicated above} }
}



\maketitle

\begin{abstract}
\textls[0]{Digital Enhanced Cordless Telecommunications 2020 New Radio (DECT-2020 NR) has garnered recognition as an alternative for cellular 5G technology in the internet of things industry. This paper presents a study centered around the analysis of the link distance performance in varying environments for DECT-2020 NR. The study extensively examines and analyzes received signal strength indicator and resulting path loss values in comparison with theoretical models, as well as packet success rates (SR) and signal-to-noise ratio against varying distances. The measurements show that with an SR of over \SI{90}{\percent}, an antenna height of \SI{1.5}{m}, indoor link distances with a single device-to-device connection with \SI{0}{dBm} transmission (TX) power can reach over \SI{60}{m} in non-line-of-sight (NLOS) areas and up to \SI{190}{m} in LOS areas with smaller \SI{-8}{dBm} TX power. Similarly, for outdoor use cases, link distances of over \SI{600}{m} can be reached with +\SI{19}{dBm} TX power.}
%The measurements show that with a success rate (SR) of over \SI{90}{\percent} and transmission (TX) power ranging from \SIrange{-20}{0}{dBm}, indoor line-of-sight (LOS) distances range from \SIrange{40}{190}{m} and for non-LOS (NLOS) use cases a distance of over \SI{60}{m} can be reached with \SI{0}{dBm} TX power. For outdoor use cases, with maximum TX power of \SI{19}{dBm}, link distances of around \SI{600}{m} can be reached with 
\end{abstract}

\vskip0.5\baselineskip

% \fi
% \IEEEpeerreviewmaketitle

%%%%%%%%%%%%%%%%%%%%%%%%%%%%%%%%%%%%%%%%%%%%%%%%%%%%%%%%%%%%%%%%%%%%%%%%%%%%%%%%%
\section{Introduction}
%%%%%%%%%%%%%%%%%%%%%%%%%%%%%%%%%%%%%%%%%%%%%%%%%%%%%%%%%%%%%%%%%%%%%%%%%%%%%%%%%
\textls[-0]{In the current era of the internet, humanity has witnessed an unprecedented level of interconnections. The growth of limited-range networks and the expansion of devices connected to these networks have led to a continuous interconnection between devices. This interconnection enables devices to collect and distribute information through a series of communication processes, often involving minimal or no human intervention \cite{Al}. These devices are embedded with intelligence and communication capabilities, including sensors, cellphones, vehicles, household devices, and medical equipment. This transformation has led to the emergence of the internet of things (IoT), marking a significant transition from the current human-focused internet to a network where devices communicate with each other autonomously \cite{Madakam}.}

Digital Enhanced Cordless Telecommunications 2020 New Radio (DECT-2020 NR) is a new wireless communication standard designed specifically for industrial IoT applications by targeting massive machine-type communications (mMTC) and/or ultra-reliable low latency communications (URLLC) capabilities. It introduces a self-organizing, decentralized mesh topology, eliminating the necessity for dedicated infrastructure like base stations or a centralized core. Within this framework, individual devices can function as intermediary hubs, facilitating multi-hop data propagation when direct device-to-device links are unfeasible \cite{Nihtilä}.

DECT-2020 NR standard has been incorporated into the 5G standards by International Telecommunication Union (ITU), with European Telecommunications Standard Institute (ETSI) publishing its inaugural version in June 2020. Notably, the ITU Radiocommunication Sector (ITU-R) has encompassed DECT-2020 NR within the 5G NR standards under International Mobile Telecommunications (IMT) 2020, or IMT-2020, technology recommendation ITU-R M.2150-2 \cite{ITU}. In line with this, the DECT-2020 NR standard, as outlined by ETSI, serves as a critical framework for achieving robust and seamless IoT device communication.

The radio interface architecture of DECT-2020 NR has four layers:  physical layer, medium access control (MAC) layer, data link Control layer (DLC) with routing service and convergence layer (CVG). Physical layer comprises with two parts: synchronization training field (STF) and data field (DF). STF enables per-packet synchronization and the gain control at the receiver. DF section of the packet encompasses two distinct channels, namely physical control channel (PCC) and physical data channel (PDC) \cite{ETSI_TS_2024_V3}.
%\textcolor{red}{The physical layer performs error detection on physical channels with indication to higher layers, forward error correction (FEC) encoding and decoding, hybrid automatic repeat request (HARQ) soft-combining, and rate matching of coded data to physical channels. It also handles the mapping of coded data onto physical channels, modulation and demodulation processes, frequency and time synchronization, and radio characteristics measurements.}
{\colorJP The PCC provides a control channel for MAC that can indicate short network ID, transmitter and receiver IDs of the transmission and used transmission parameters of the PDC. Transmission parameters include transmission power, modulation and coding scheme (MCS), transmission length, number of spatial streams, and redundancy versions. Furthermore, PCC encompasses hybrid automatic repeat request (HARQ) feedback channel quality indicator (CQI), and codebook index for MIMO transmissions. The PDC then contains the MAC PDU, containing the MAC protocol data and control units as defined in \cite{ETSI_TS_2024_V3}.}

In this paper, radio propagation models are first presented and an extensive measurement campaign is performed to compare the achieved link distances with theoretical ones. The aim is to show how well DECT-2020 NR point-to-point connections, or in mesh network terms one-hop connections, are able to perform in numerous different environments with good packet success rates. These environments include multiple indoor and outdoor locations with both line-of-sight (LOS) and non-line-of-sight (NLOS) connections.
%%%%%%%%%%%%%%%%%%%%%%%%%%%%%%%%%%%%%%%%%%%%%%%%%%%%%%%%%%%%%%%%%%%%%%%%%%%%%%%%%
%\section{Radio Wave Propagation Principle and Modeling}
\section{Radio Propagation Models}
\subsection{Propagation Models}
%%%%%%%%%%%%%%%%%%%%%%%%%%%%%%%
Radio wave propagation models, which typically depict the radio environment as a function of frequency and distance between the terminals, are utilized to predict the path losses of DECT-2020 NR link level connections. The models considered for comparison with the measurements are the following ones:

%%%%%%%%%%%%
\subsubsection{Free Space Path Loss Model}
%%%%%%%%%%%%
The basic and most well-known model for the attenuation of the transmitted signal's power between two antennas through free space with a direct line of sight (LOS) between the transmitter and receiver is called the free space path loss (FSPL). The FSPL equation can be written as \cite{FSPL}:
\begin{equation}
\label{eq:1}
\mathit{FSPL}~(\text{dB}) = 20 \log_{10}(d) + 20 \log_{10}(f) + 20 \log_{10}\left( \frac{4 \pi}{c} \right),
\end{equation}
where $d$ is the distance between antennas (m), $f$ is frequency of the signal (Hz), and $c$ is speed of light (m/s).% (approximately \( 3 \times 10^8 \)~m/s).

%%%%%%%%%%%%
\subsubsection{3GPP TR 38.901 Path Loss Model}
%%%%%%%%%%%%
3GPP technical report version 38.901 focuses on frequencies ranging from \SIrange{0.5}{100}{GHz}. In this model, indoor office environments and shopping malls are referred to as InH, indoor factories containing heavy machinery, assembly lines, and varying densities of clutter are referred to as InF. The two most relevant equations for this study from this model are the following \cite{3GPPTR38.901}: 
\begin{equation}
\label{eq:2}
\mathit{PL}_{\text{InH-LOS}} (\text{dB}) = 32.4 + 17.3 \log(d) + 20 \log(f_\text{c}), \text{and}
\end{equation}
\begin{equation}
\label{eq:3}
\mathit{PL}_{\text{InF-LOS}} (\text{dB}) = 31.84 + 21.50 \log(d) + 19 \log(f_\text{c}),
\end{equation}
where \( d \) is defined as the distance between the antenna and the measurement point (m) and $f_\text{c}$ is the carrier frequency (GHz).

%%%%%%%%%%%%
\subsubsection{Two-Ray Ground Reflection Model}
%%%%%%%%%%%%
Two-ray ground-reflection model accounts for both the direct signal path and the signal reflected off the ground. The equation stands as below \cite{Rappaport}:
\begin{equation}
\label{eq:4}
\mathit{PL}~(\text{dB}) = 40 \log_{10}(d) - 10 \log_{10}(G h_\text{t}^2 h_\text{r}^2),
\end{equation}
% \( P_\text{t} \) is transmitting power, \( P_\text{r} \) is receiving power, 
where \( d \) is the distance between the transmitter (TX) and receiver (RX), \( h_\text{t} \) is the height of the TX, \( h_\text{r} \) is the height of the RX, and
\( G \) is the combined TX-RX antenna gain. %, \( G_\text{t} \cdot G_\text{r} \). 

%%%%%%%%%%%%
\subsubsection{Okumura-Hata Model}
%%%%%%%%%%%%
The Okumura-Hata model is particularly well-suited for urban environments with numerous buildings with relatively few tall obstructions. The model is formulated as follows\cite{hata1980empirical}:
\begin{equation}
\label{eq:5}
\begin{split}
L_{\text{U}} (\text{dB}) = 69.55 + 26.16 \log_{10} (f) - 13.82 \log_{10} (h_{\text{B}}) - C_{\text{H}} \\
    + [44.9 - 6.55 \log_{10}(h_{\text{B}})] \log_{10}(d). 
\end{split}
\end{equation}
where \(L_{\text{U}} \) is PL in urban areas (dB), \( f \) is the utilized frequency (MHz), \(h_{\text{B}} \) is the height of the base station antenna (m), \(C_{\text{H}}\) is antenna height correction factor, and \( d \) is the distance between the base station and mobile station (km).
%%%%%%%%%%%%
\subsubsection{COST-231 Hata Model}
%%%%%%%%%%%%
The COST-231 Hata model is applicable in the frequency range of \SIrange{500}{2000}{MHz}. The basic equation for PL in dB is given by \cite{damosso1999digital}:
\begin{equation}
\label{eq:8}
\begin{split}
\mathit{PL}~(\text{dB}) = 46.3 + 33.9\log_{10}(f) - 13.82\log_{10}(h_{\text{b}}) - \alpha h_{\text{m}} + \\
[44.9 - 6.55\log_{10}(h_{\text{b}})]\log_{10}(d) + C_m, 
\end{split}
\end{equation}
where \( f \) is the utilized frequency (MHz), \( d \) is the distance between access points (AP) and customer premises equipment (CPE) antennas (km), \( h_{\text{b}} \) is the AP antenna height above ground level (m), \( h_{\text{m}} \) is the CPE antenna height above ground level (m), $\alpha h_{\text{m}}$ is a city size dependent variable, and \( C_{\text{m}} \) is 0~dB for suburban or open environments and 3~dB for urban environments. 
%%%%%%%%%%%%%%%%%%%%%%%%%%%%%%%
\subsection{Empirical Path Loss}
%%%%%%%%%%%%%%%%%%%%%%%%%%%%%%%
Empirical PLs were calculated using the formula below to compare with the theoretical PL values for the corresponding scenarios. The empirical PL formula is defined as:
\begin{equation}
\label{eq:11}
\mathit{PL}_{\text{emp}}~(\text{dB}) = P_{\text{tx}} - P_{\text{rx}} + G_{\text{tx}} - L_{\text{tx}}+ G_{\text{rx}} - L_{\text{rx}},
\end{equation}
where $P_{\text{tx}}$ is the transmission power of TX (dBm), $P_{\text{rx}}$ is the receiving power of RX (dBm), $G_{\text{tx}}$ is the TX antenna gain (dB), $L_{\text{tx}}$ is the TX side connector losses (dB), $G_{\text{rx}}$ is the RX antenna gain (dB), and $L_{\text{rx}}$ is the RX side connector losses (dB).
%%%%%%%%%%%%%%%%%%%%%%%%%%%%%%%%%%%%%%%%%%%%%%%%%%%%%%%%%%%%%%%%%%%%%%%%%%%%%%%%%
\section{Data Acquisition and Assessment}
%%%%%%%%%%%%%%%%%%%%%%%%%%%%%%%%%%%%%%%%%%%%%%%%%%%%%%%%%%%%%%%%%%%%%%%%%%%%%%%%%

%%%%%%%%%%%%%%%%%%%%%%%%%%%%%%%
\subsection{Equipment Details}
\begin{figure}[h]
\subfloat[]{\includegraphics[height=1.8 in]{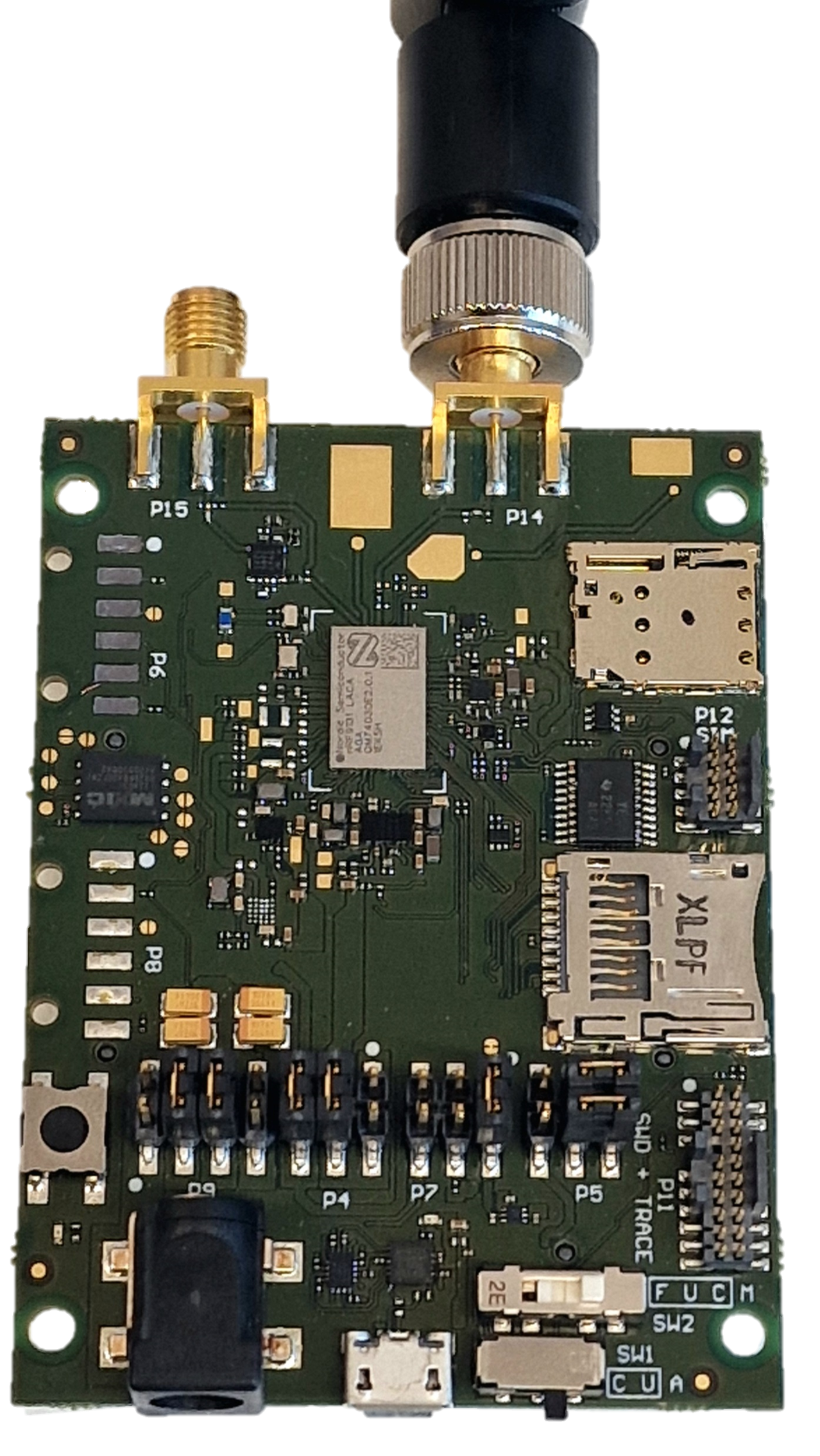}}
\subfloat[]{\includegraphics[height=1.8 in]{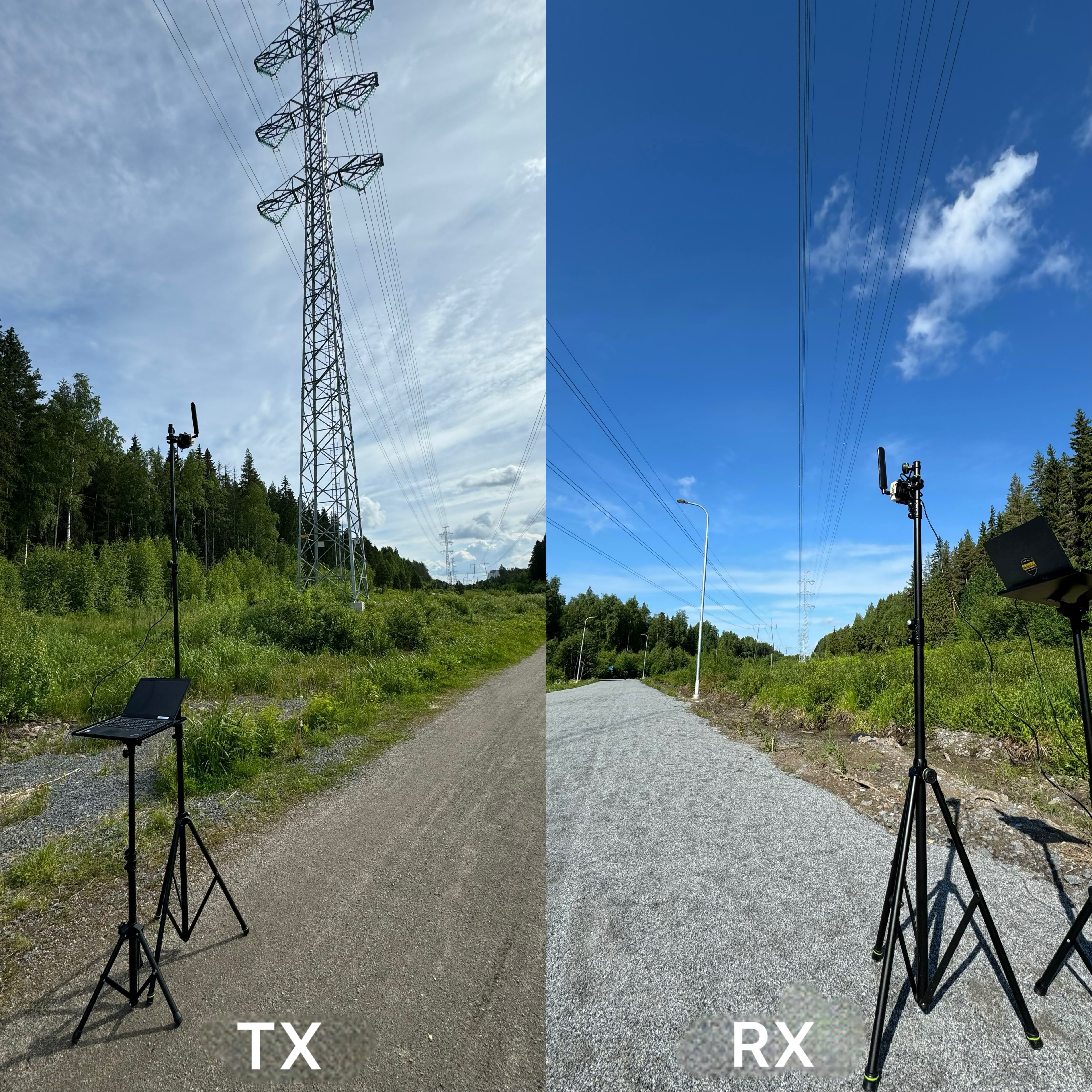}}
\centering
\caption{a) The utilized nRF9131 board and b) Hallila power line measurement scenario.}
\label{fig:hallila}
\end{figure}

%%%%%%%%%%%%%%%%%%%%%%%%%%%%%%%
For analyzing the propagation effects, pre-commercial devices have been utilized. These devices, model nRF9131 as shown in Fig.~\ref{fig:hallila}, were manufactured by Nordic Semiconductor and support the new DECT-2020 NR standard \cite{Nordic}. Nordic Semiconductor's proprietary software was employed to command and control these devices. Once the devices were set up for a link distance measurement, data was collected and stored for post-processing. 
%using Mobaxterm software. Although the devices were primarily tethered to continuous power from laptops during measurements, the nRF9131 can also be operated in a standalone battery mode for limited periods. 
%Pulse Larsen blade antennas were used at both TX or client and RX or server devices. 
% and connector type is W5095K SMA male connector.

%Wirepas Oy, a Finnish company that mainly provides enterprise IoT solutions, have developed Wirepas 5G Mesh, which is the first non-cellular 5G network. It is capable of connecting millions of devices, such as the nRF9131, even in challenging environments, and operates at a frequency of \SI{1.9}{GHz} \cite{WirepasMesh}. DECT-2020 NR radio interface with mesh architecture, developed by Wirepas utilizing devices like nRF9131, is applied in diverse use cases, including electricity network quality of service monitoring to support the transition from fossil fuels to distributed renewable energy, and condition monitoring in industrial facilities to enhance machine monitoring \cite{USWA}.

%%%%%%%%%%%%%%%%%%%%%%%%%%%%%%%
\subsection{Measurement Setup}
%%%%%%%%%%%%%%%%%%%%%%%%%%%%%%%
A point-to-point connection was employed for measuring average received signal strength indicator (RSSI), path loss (PL), signal-to-noise ratio (SNR) and success rate (SR) by collecting 300 measurement samples (one sample per second) at each location point. Both the TX and RX were mounted on separate tripods to achieve stationary positions and desired heights. Moreover, antennas were connected to both of the devices as shown in Fig.~\ref{fig:hallila}. These Pulselarsen W5095X antennas are vertically polarized, omnidirectional blade antennas.%and have a frequency depended (from \SIrange{700}{2200}{MHz}) gain \cite{Nordic}.

For all indoor distance measurements, a laser distance meter was employed to precisely measure the distance between the TX and RX devices. In outdoor scenarios, online mapping tools were utilized to determine the distances between the devices by utilizing GPS coordinates measured at both ends with a common smart phone. 

RSSI data from the RX, along with cyclic redundancy check (CRC) success and failure counts, were independently recorded and stored. Lastly, MATLAB software was employed to process the data from these files.

%%%%%%%%%%%%%%%%%%%%%%%%%%%%%%%
\subsection {Data Assessment Procedure}
%%%%%%%%%%%%%%%%%%%%%%%%%%%%%%%

%%%%%%%%%%%%
\subsubsection{Received Signal Strength Indicator and Signal to Noise Ratio Calculation}
%%%%%%%%%%%%
Following the acquisition of each dataset, the RSSI levels for both PCC and PDC were systematically organized. MATLAB script was subsequently employed to compute key statistical parameters, including the mean, standard deviation, maximum, and minimum values. %\textcolor{blue}{The analysis revealed low RSSI variability across locations, with standard deviations (SD) ranging from 0.34 dB to 4.17 dB} 
The RSSI and SNR measurements, initially recorded in logarithmic scales, underwent a conversion process to linear scale to facilitate average value calculations, before being reconverted to logarithmic scale. This methodical approach was applied to the raw data (RSSI and SNR data sets) for each measured distance, ensuring a comprehensive representation of signal characteristics across varying spatial parameters. Furthermore, this average RSSI value in logarithmic form is subtracted from the transmitted power, while internal and connector losses along with TX-RX gains (total 1 dB per side) are added to compute the empirical PLs, according to (\ref{eq:11}). The measurements were conducted using omnidirectional antennas, which closely align with the assumptions made in the propagation models considered in this study.

%%%%%%%%%%%%
\subsubsection{Success Rate Analysis}
%%%%%%%%%%%%
In the measurement process, the device utilized a continuous reception method without implementing any re-transmission mechanism for packet error correction. Instead, CRC was employed as the primary means of error detection. The `CRC Okay Count' served as central metric, enabling the quantification of successfully received packets that passed the error detection process. SRs were quantified as percentages, utilizing formulas below:
\begin{equation}
\label{eq:12}
\mathit{SR}_{\text{PCC}} = \left( \frac{\text{PCC CRC Okay Count}}{\text{Request Count}} \right) \times 100 
\end{equation}
\label{eq:13}
\begin{equation}
\mathit{SR}_{\text{PDC}} = \left( \frac{\text{PDC CRC Okay Count}}{\text{Request Count}} \right) \times 100 
\end{equation}

Finally, RSSI values, both theoretical and empirical PL equations and SRs were visualized through distinct curves plotted against distance. This meticulous data collection and analysis approach provided in-depth insights into the performance characteristics and reliability of the studied devices, while also offering valuable information about their applicability in various indoor and outdoor scenarios. 

%%%%%%%%%%%%
\subsection{Data Collection from Measurement Scenarios}
%%%%%%%%%%%%
Link distance measurements were carried out at various sites situated at Tampere, Finland, to collect the data. These sites can be categorized as indoor and outdoor locations. Indoor measurements were conducted at Tampere University Hervanta campus (buildings ``Konetalo" and ``Rakennustalo") and Valmet Oy industry premises. Outdoor locations included various sites and areas around Tampere University, Ahvenisjärvi, Opiskelijankatu, Hallila, Kangasala, and Kaukajärvi, encompassing urban, sidewalk, forested, and lakeside environments. The general and DECT-2020 NR specific parameters considered for the measurements are shared in Table~\ref{table_parameters}.
\begin{table}[h]
\renewcommand{\arraystretch}{1.1}
\caption{Considered parameters for DECT-2020 NR link-level measurements}
\label{table_parameters}
\centering
\scalebox{0.9}{
\begin{tabular}{c|c}
\hline
\textbf{Parameter} & \textbf{Value} \\
\hline
Payload length (in slots) & 4 \\
\hline
Modulation and Coding Scheme (MCS) & 1 \& 4 \\
\hline
Successive Packet Transmission Interval & \SI{1}{s} \\
\hline
DECT-2020 Channel & 1677 \& 1667 \\
\hline
TX Power & \SI{-20}{dBm}, \SI{0}{dBm}, +\SI{19}{dBm} \\
\hline
Node Height above Ground Level & \SI{1.5}{m} \\
\hline
Antenna Gain & \SI{1.07}{dB} $\pm$ \SI{1}{dB} \\
\hline
%Receiver Sensitivity Threshold & \SI{-108}{dBm} \\
%\hline
Carrier Frequency & \SI{1899}{MHz} \\
\hline
Channel Bandwidth & \SI{1.728}{MHz}\\
\hline
\end{tabular}}
\end{table}
%%%%%%%%%%%%%%%%%%%%%%%%%%%%%%%%%%%%%%%%%%%%%%%%%%%%%%%%%%%%%%%%%%%%%%%%%%%%%%%%%
\section{Results and Analysis}
%%%%%%%%%%%%%%%%%%%%%%%%%%%%%%%%%%%%%%%%%%%%%%%%%%%%%%%%%%%%%%%%%%%%%%%%%%%%%%%%%
To ensure reliable communication and satisfactory performance of the devices under examination, the SR is required to exceed \SI{90}{\percent}. This is considered a good threshold for a well-performing single-link connection when considering only one-hop and targeting maximum distance, {\colorJP without introducing an extensive amount of re-transmission in the link. In fact, \SI{10}{\percent} packet error rate is the typical assumption in wireless data networks such as Long Term Evolution (LTE) \cite{3GPPTR36.213}.} {\colorJP Although a typical DECT-2020 NR mesh network is expected to consist of a large number of nodes enabling longer end-to-end distances with multiple hops, it can still guarantee up to \SI{100}{\percent} SR by enabling link-specific re-transmission using HARQ for any packet error occurring in a link.}

In order to help to analyse and visualize the results, nonlinear least squares (NLS) curve fitting was also applied to the measurement data for logarithmic scale PL values versus distance graphs. This was utilized, when several distances were measured in one location. 
%%%%%%%%%%%%%%%%%%%%%%%%%%%%%%%
\subsection{Minimum Transmission Power Requirements}
%%%%%%%%%%%%%%%%%%%%%%%%%%%%%%%
A crucial parameter for long-term device deployment is minimizing TX power while maximizing range, which optimizes battery life. Additionally, utilizing optimum TX power reduces interference in the spectrum, improving system capacity and co-existence with other systems \cite{LPWA}. To elucidate these power-distance relationships across various propagation scenarios, Table~\ref{indoor} and Table~\ref{outdoor} present a comprehensive summary of the minimum power requirements necessary to achieve maximum communication distances under the condition of reliable communication.
\begin{table}[h]
\renewcommand{\arraystretch}{1.0}
\centering
\caption{Minimum transmission power requirements for reliable communication at maximum distance in \underline{indoor} scenarios}
\label{indoor}
\scalebox{0.55}{
\begin{tabular}{|c|c|c|c|c|}
\hline
\shortstack{Location} & \shortstack{Propagation Type} & \shortstack{Devices Located at / \\Separated By} & \shortstack{Maximum Distance \\ Measured (m)} & 
\shortstack{Minimum TX Power\\ Required (dBm)} \\
\hline
Konetalo 2nd Floor (1) & LOS & Straight Corridor & 40 & \textminus20 \\
Rakennustalo 2nd  Floor (2) & LOS & Straight Main Corridor & 120 & 0 \\
Valmet Office (3) & LOS & Straight Main Corridor & 137 & 0 \\
Valmet Factory (4) & LOS & Long Factory Area  & 190 & \textminus8 \\
\hline
Konetalo Inter-floors & NLOS & Concrete Floors (three floors) & 10.71 & 0 \\
Valmet Office & NLOS & Brick Wall \& Glass & 61 & 0 \\
Valmet Office & NLOS & Brick Wall, Glass and Cubicles & 61 & 0 \\
Valmet Office & NLOS & Circular Main Corridor & 111 & +19 \\
\hline
\end{tabular}}
\end{table}

\begin{table}[h]
\renewcommand{\arraystretch}{1.0}
\centering
\caption{Minimum transmission power requirements for reliable communication at maximum distance in \underline{outdoor} scenarios}
\label{outdoor}
\scalebox{0.55}{
\begin{tabular}{|c|c|c|c|c|}
\hline
\shortstack{Location} & \shortstack{Propagation\\Type} & \shortstack{Devices Located at /\\Separated By} & \shortstack{Maximum Distance\\Measured (m)} & \shortstack{Minimum TX Power\\Required (dBm)} \\
\hline
Kampusareena and Main Building Outdoors (5) & LOS & Free-Space & 40 & \textminus20 \\
Ahvenisjärvi Football Ground (6) & LOS & Free-Space & 205 & \textminus8 \\
Ahvenisjärvi Forest (7) & NLOS & Forest and Lake & 265 & +16 \\
Opiskelijankatu Sidewalk (8) & LOS & Walking Bridge & 610 & +19 \\
Hallila Power Line (9) & LOS & Free-Space & 650 & +19 \\
Kaukajärvi Lake & LOS& Lake & 2294& +19\\
Kangasala Observation Tower & LOS & Terrain & 2470 & +19 \\
\hline
\end{tabular}}
\end{table}

%%%%%%%%%%%%%%%%%%%%%%%%%%%%%%%
\subsection{PL versus Distance}
\begin{figure}[h]
\centering
    % Subfigure for PCC
\includegraphics[width=0.9\linewidth, trim=40 210 40 210, clip]{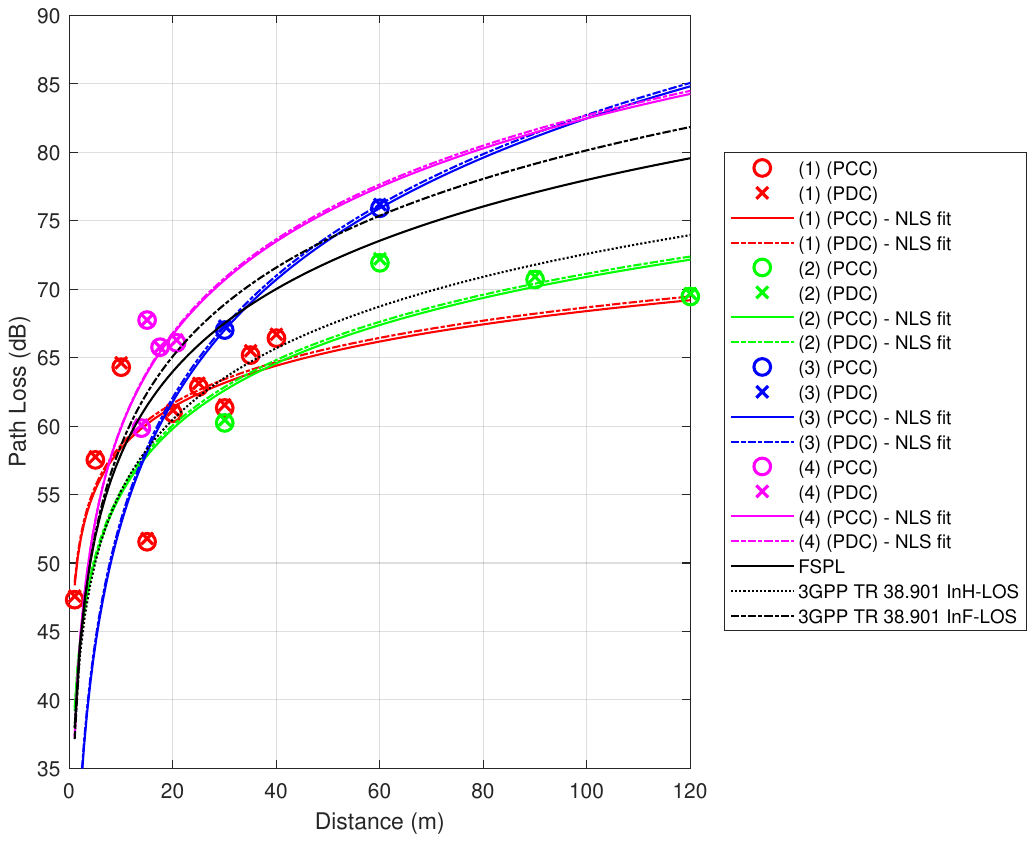}
\vspace{-4mm}
\caption{PL versus distance in indoor scenarios for TX = \SI{-20}{dBm}. Site labels correspond to Table II locations. RSSI standard deviation: 0.34–4.17 dB.}
\label{PL_Indoor_-20dBm}
\end{figure}

%%%%%%%%%%%%%%%%%%%%%%%%%%%%%%%
\textls[-20]{Fig.~\ref{PL_Indoor_-20dBm} illustrates empirical PL against distance for various indoor corridor LOS scenarios. Konetalo 2nd floor corridor consistently has the lowest PL, staying below 3GPP TR 38.901 InH-LOS threshold. Valmet factory corridor shows highest PL among all, but exhibits a moderate PL compared to InF-LOS model.} 

\begin{figure}[h]
\centering
\includegraphics[width=0.9\linewidth, trim=40 210 40 210, clip]{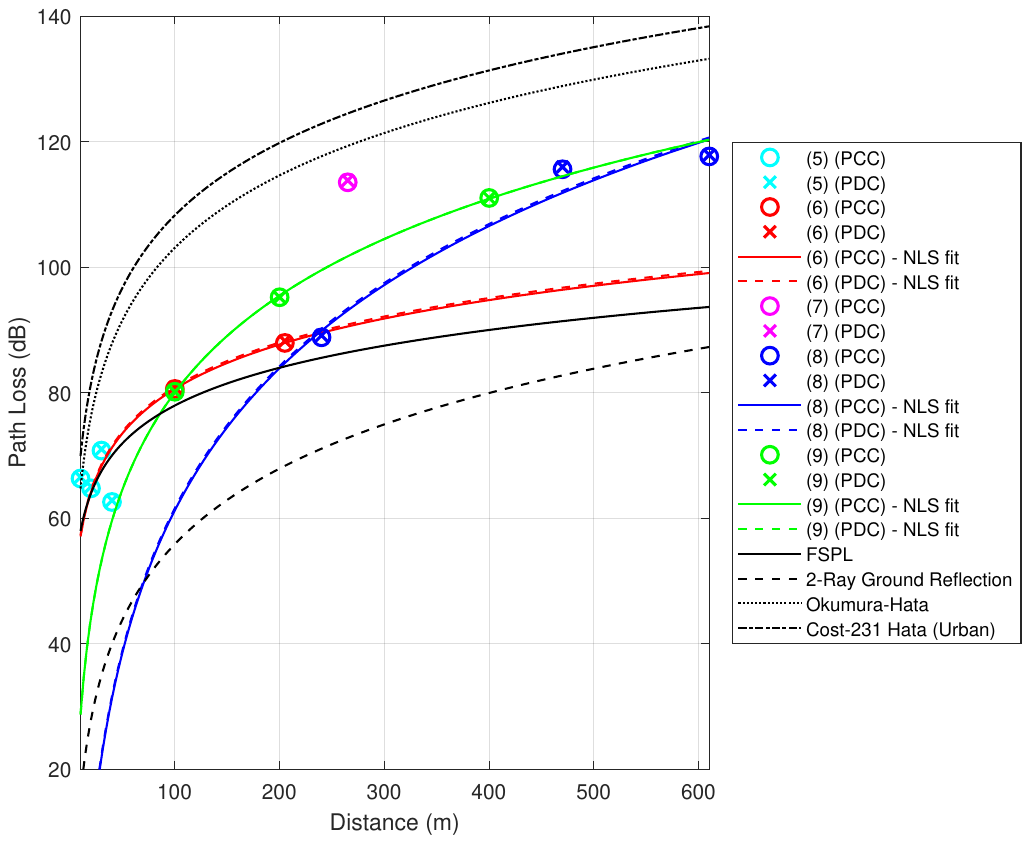}
\vspace{-4mm}
\caption{PL versus distance in outdoor scenarios for TX = +\SI{19}{dBm}. Site labels correspond to Table III locations. RSSI standard deviation: 0.59–2.75 dB.}
\label{PL_Outdoor_19dBm}
\end{figure}

\textls[-20]Moreover, Fig.~\ref{PL_Outdoor_19dBm} illustrates PL and distance relationship for various environments at outdoors LOS for \SI{+19}{dBm} TX power. The heights of the TX and RX have been designated as \SI{10}{m} and \SI{1.5}{m} respectively for the theoretical models. Ahvenisjärvi football ground showed a very close performance with the FSPL curve, whereas Opiskelijankatu sidewalk exhibited one of the highest PLs indicating impact of multipath propagation due to pedestrians and live vehicular traffic as well as more vegetation along the road. Kampusareena-Mainbuilding outdoor scenario showed notable variation in PL values as the devices are located between two buildings and Kampusareena is metal coated. Therefore, the signal got reflected very well from the buildings and showed tunneling effect in guiding the signal well between the closely located buildings. Under Hallila power lines, the link distance is very similar to the Opiskelijankatu sidewalk case.
%might be an impacting factor for DECT-2020 NR transmissions. 
Ahvenisjärvi forest PL remained highest in comparing with other scenarios of the same distance due to the presence of a thick forest. However, all measured scenarios consistently showed lower PL than the Okumura-Hata and COST-231 Hata models predict at all distances and higher PL than FSPL model predicts after \SI{100}{m} distance. %In all cases, PL values were observed to be increasing with the increment of distance, aligning the concept with theoretical PL models. 

The outdoor LOS measurements were performed with \SI{+19}{dBm} TX power to assess device performance at maximum distances with different heights, as shown in Table~\ref{max distance_pl_vs_distance}. Kangasala observation tower outperformed other sites, but this is due to its considerable height advantage (TX-RX height difference is \SI{52.6}{m}). Excluding height differences, Kaukajärvi lake showed favorable propagation conditions over water, as one might expect. All measured scenarios exhibited significantly lower PL than the two models Okumura-Hata and COST-231 Hata predicted, which is due to the more pessimistic nature of these models.  % Despite the varying heights and locations, the measured PL values remained consistently clustered between \SIrange{105}{120}{dB} across the distance range, indicating consistent performance in different LOS outdoor scenarios.
\begin{table}[h!]
\renewcommand{\arraystretch}{1.0}
\centering
\caption{PL for different (maximum) link distances at outdoors.}
\label{max distance_pl_vs_distance}
\centering
\scalebox{0.6}{
\begin{tabular}{|c|c|c|c|c|c|c|c|c|}
\hline
\shortstack{Scenario} & \shortstack{Distance \\(m)} & \shortstack{Height\\ Difference \\(m)} & \shortstack{Empirical\\PL  PCC\\ (dB)} & \shortstack{Empirical\\PL  PDC\\ (dB)} & \shortstack{FSPL\\(dB)} & \shortstack{2-Ray\\Model PL \\(dB)} & \shortstack{Okumura \\Hata\\Model\\PL (dB)} & \shortstack{COST-231\\Hata Model\\PL(dB)} \\
\hline
Hallila Power Line & 650 & 11.58 & 110.56 & 110.84 & 99.55 & 91.63 & 139.44 & 145.60 \\
\hline
Kaukajarvi Lake & 2294 & 1 & 120.38 & 120.61 & 105.16 & 115.00 & - & - \\
\hline
Kangasala Tower & 2470 & 52.6 & 113.70 & 113.96 & 105.83 & 96.93 & 144.50 & 149.67 \\
\hline
\end{tabular}}
\end{table}
%%%%%%%%%%%%%%%%%%%%%%%%%%%%%%%
\subsection{SR \& SNR versus RSSI}
\begin{figure}[h]
\centering
    % Subfigure for PCC
\includegraphics[width=0.9\columnwidth]{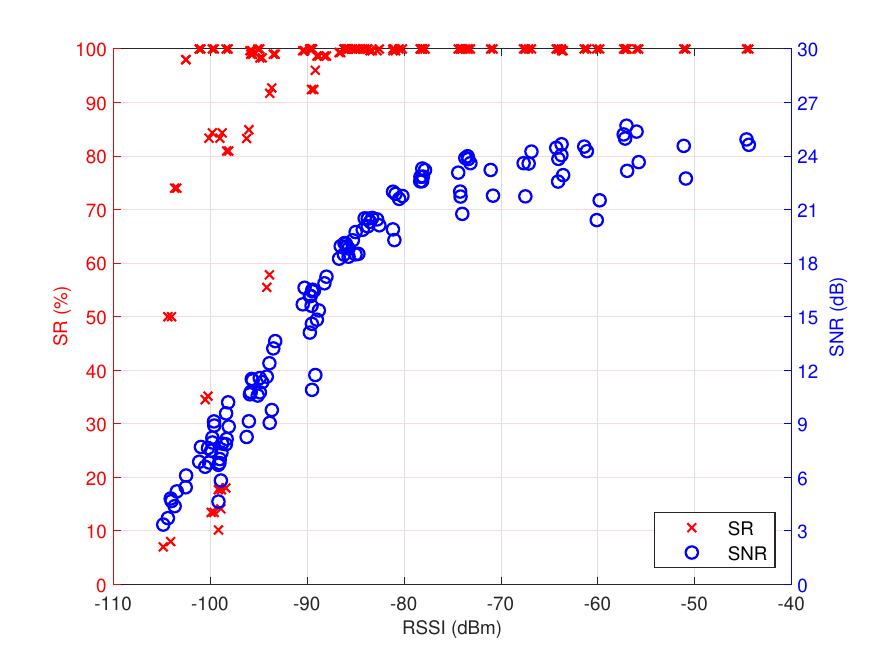}
\caption{SR and SNR as a function of RSSI at all indoor scenarios.}
\label{SR_SNR_RSSI_Indoors}
\end{figure}

\begin{figure}[h]
\centering
    % Subfigure for PCC
\includegraphics[width=0.9\columnwidth]{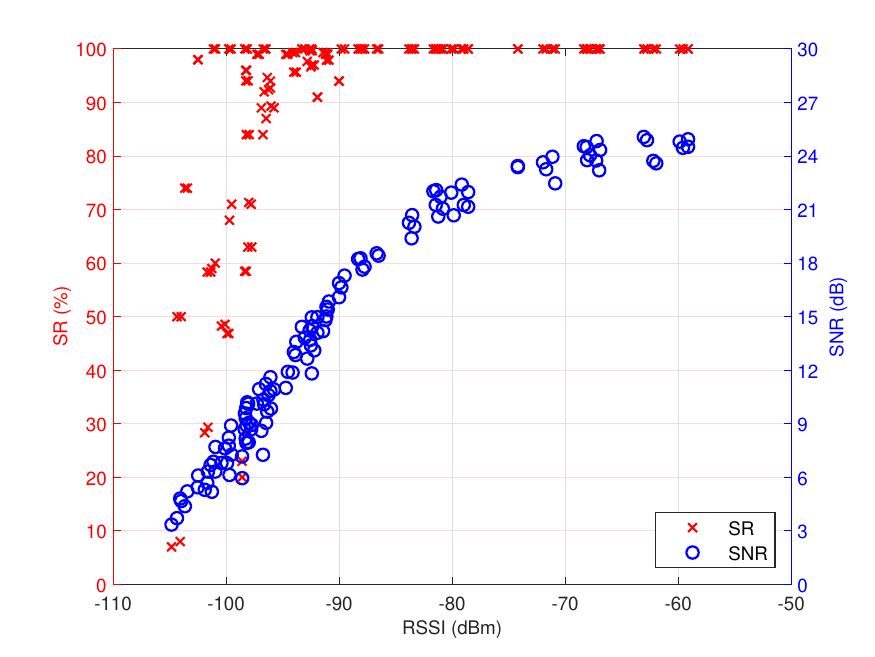}
\caption{SR and SNR as a function of RSSI at all outdoor scenarios.}
\label{SR_SNR_RSSI_Outdoors}
\end{figure}

%%%%%%%%%%%%%%%%%%%%%%%%%%%%%%%
In this section, SR and SNR are simultaneously analyzed as a function of RSSI. From Fig.~\ref{SR_SNR_RSSI_Indoors} and Fig.~\ref{SR_SNR_RSSI_Outdoors}, it can be summarised that across all indoor and outdoor environments, SR consistently exceeded \SI{90}{\percent} when RSSI values remained above \SI{-90}{dBm} for indoors and above \SI{-95}{dBm} for outdoors, demonstrating the capability to perform well even with weaker signal power levels.
\vspace{-0.25em}

The comparison of SNR to RSSI shows a more gradual increase compared to the sharp rise in the case of SR versus RSSI. Generally, SRs surpassed \SI{90}{\percent} when SNR values reached around \SIrange{11}{12}{dB} for indoors and around \SIrange{12}{15}{dB} for outdoors. SR data points cluster heavily near \SI{100}{\percent} for stronger RSSI values, indicating excellent performance, while SNR values spread more, especially at stronger RSSI levels.
%%%%%%%%%%%%%%%%%%%%%%%%%%%%%%%%
%\subsection{Reliability of the Analysis}
%%%%%%%%%%%%%%%%%%%%%%%%%%%%%%%%
%%Data reliability was assessed through RSSI analysis. 
%For all locations (a total of 18 measurement places), 300 samples were collected and the average values were used for the analysis. Standard deviations ranged from 0.5 to 5.5. Distances were measured utilizing separate GPS coordinates from a typical smart phone for outdoor locations and indoor locations .

%%%%%%%%%%%%%%%%%%%%%%%%%%%%%%%%%%%%%%%%%%%%%%%%%%%%%%%%%%%%%%%%%%%%%%%%%%%%%%%%%
\section{Conclusions}
%%%%%%%%%%%%%%%%%%%%%%%%%%%%%%%%%%%%%%%%%%%%%%%%%%%%%%%%%%%%%%%%%%%%%%%%%%%%%%%%%
This paper presented a measurement-based study for practical DECT-2020 NR link distances in indoor and outdoor scenarios with pre-commercial equipment. %The analysis provided insights such as signals weakening due to distance and other predictable and unpredictable factors.
%Moreover, m
The maximum one-hop link distance communication ranges were determined for diverse outdoor settings, including open fields, forests, urban areas, terrains and over water surfaces. 

The measurements show that with an SR of over \SI{90}{\percent} and TX power ranging from \SIrange{-20}{0}{dBm}, the indoor LOS distances range from \SIrange{40}{190}{m} and for NLOS use cases a distance of over \SI{60}{m} can be reached with \SI{0}{dBm} TX power. For outdoor use cases, with the maximum TX power of \SI{19}{dBm}, link distances of around \SI{600}{m} can be reached in LOS environment with low device heights. With deploying devices higher e.g. light poles, roofs or masts, significantly longer distances can be achieved as demonstrated by measurements done at Kangasala observation tower.

The indoor scenarios PL characteristics predominantly aligned with the 3GPP TR 38.901 InH-LOS model, which can be stated as the most suitable model for the measurements. The predominant deployment of devices at similar heights in outdoors indicates the need for a more sophisticated theoretical model to be considered as the most suitable model.

With the achieved link distance results, it can be concluded that indoor networks can be realized with small TX powers enabling, for example, battery operated networks to function for a long time. Furthermore, large outdoor areas can be covered with suitable link distances, that is, having up to several hundred meters between the nodes, when node installation height is only \SI{1.5}{m}, {\colorJP enabling reduction of network installation and maintenance cost compared to high mass deployments typically used in LPWAN or cellular technologies.}

The results of this study advance both device performance under DECT-2020 NR communication protocols while providing insights into IoT-enabled systems' propagation characteristics. The findings not only show the devices’ readiness for industrial deployment in smart metering, electricity network QoS monitoring and industrial facility condition monitoring, as well as smart building and smart city applications, but also establish benchmarks for evaluating performance across various environmental conditions.
% Future research may utilize these results to optimize antenna configurations, develop link-level simulators, and evaluate performance under varying environmental and hardware constraints.
%The performance of PCC and PDC was  similar. %The limitations of this study include incomplete consideration of Aluminum structures in industrial scenarios, lack of investigation into high-vibration equipment effects, and absence of long-term performance data for all scenarios.

%In future, there is a scope of conducting a performance comparison between DECT-2020 NR-based IoT devices and various IoT technologies across multiple environments would be interesting to analyze. Moreover, evaluating the performance of DECT-2020 NR under high device mobility conditions, comparing it with cellular technologies will address potential challenges in dynamic environments and assess its suitability for mobile IoT applications. 
%Future studies will be focusing on evaluating the performance of DECT-2020 NR under high mobility use cases.

%%%%%%%%%%%%%%%%%%%%%%%%%%%%%%%%%%%%%%%%%%%%%%%%%%%%%%%%%%%%%%%%%%%%%%%%%%%%%%%%%
\section*{Acknowledgment}
%%%%%%%%%%%%%%%%%%%%%%%%%%%%%%%%%%%%%%%%%%%%%%%%%%%%%%%%%%%%%%%%%%%%%%%%%%%%%%%%%
The work has been supported by the CELTIC-NEXT "Ultra Scalable Wireless Access" (USWA) project.

%%%%%%%%%%%%%%%%%%%%%%%%%%%%%%%%%%%%%%%%%%%%%%%%%%%%%%%%%%%%%%%%%%%%%%%%%%%%%%%%%%%%%%%%%%%%%%%%%%%%%%%%%

%\balance
\bibliographystyle{IEEEtran}
\bibliography{bibliography}

\begin{thebibliography}{10}
\providecommand{\url}[1]{#1}
\csname url@samestyle\endcsname
\providecommand{\newblock}{\relax}
\providecommand{\bibinfo}[2]{#2}
\providecommand{\BIBentrySTDinterwordspacing}{\spaceskip=0pt\relax}
\providecommand{\BIBentryALTinterwordstretchfactor}{4}
\providecommand{\BIBentryALTinterwordspacing}{\spaceskip=\fontdimen2\font plus
\BIBentryALTinterwordstretchfactor\fontdimen3\font minus \fontdimen4\font\relax}
\providecommand{\BIBforeignlanguage}[2]{{%
\expandafter\ifx\csname l@#1\endcsname\relax
\typeout{** WARNING: IEEEtran.bst: No hyphenation pattern has been}%
\typeout{** loaded for the language `#1'. Using the pattern for}%
\typeout{** the default language instead.}%
\else
\language=\csname l@#1\endcsname
\fi
#2}}
\providecommand{\BIBdecl}{\relax}
\BIBdecl

\bibitem{Al}
A.~Al-Fuqaha, M.~Guizani, M.~Mohammadi, M.~Aledhari, and M.~Ayyash, ``{Internet of Things: A Survey on Enabling Technologies, Protocols, and Applications},'' \emph{IEEE Communications Surveys \& Tutorials}, vol.~17, no.~4, pp. 2347--2376, 2015.

\bibitem{Madakam}
S.~Madakam, R.~Ramaswamy, and S.~Tripathi, ``{Internet of Things (IoT): A Literature Review},'' \emph{Journal of Computer and Communications}, vol.~3, pp. 164--173, 04 2015.

\bibitem{Nihtilä}
T.~Nihtilä and H.~Berg, ``{Energy Consumption of DECT-2020 NR Mesh Networks},'' in \emph{2022 Joint European Conference on Networks and Communications \& 6G Summit (EuCNC/6G Summit)}, 2022, pp. 196--201.

\bibitem{ITU}
{ITU}, ``{Detailed specifications of the terrestrial radio interfaces of International Mobile Telecommunications-2020 (IMT-2020)},'' {ITU}, {ITU-R M.2150-2}, December 2023.

\bibitem{ETSI_TS_2024_V3}
{ETSI}, ``{DECT-2020 New Radio (NR); Part 3: Physical layer; Release 1},'' {ETSI}, {ETSI TS 103 636-3}, March 2024.

\bibitem{FSPL}
S.~Cerwin, \emph{{Radio Propagation and Antennas: A Non-Mathematical Treatment of Radio and Antennas}}.\hskip 1em plus 0.5em minus 0.4em\relax AuthorHouse, 2019.

\bibitem{3GPPTR38.901}
{3GPP}, ``{Study on channel model for frequencies from 0.5 to 100 GHz},'' {3GPP}, {3GPP TR 38.901}, November 2020.

\bibitem{Rappaport}
T.~S. Rappaport, \emph{\BIBforeignlanguage{eng}{Wireless Communications Principles and Practice, Second Edition}}.\hskip 1em plus 0.5em minus 0.4em\relax Pearson, 2001.

\bibitem{hata1980empirical}
M.~Hata, ``Empirical formula for propagation loss in land mobile radio services,'' \emph{IEEE transactions on Vehicular Technology}, vol.~29, no.~3, pp. 317--325, 1980.

\bibitem{damosso1999digital}
E.~Damosso, L.~M. Correia \emph{et~al.}, ``Digital mobile radio towards future generation systems,'' \emph{COST 231 final report}, vol.~13, 1999.

\bibitem{Nordic}
\BIBentryALTinterwordspacing
{nRF9131}. [Online]. Available: \url{www.nordicsemi.com/Products/nRF9131}
\BIBentrySTDinterwordspacing

\bibitem{3GPPTR36.213}
{3GPP}, ``{Evolved Universal Terrestrial Radio Access (E-UTRA)},'' {3GPP}, {3GPP TR 36.213}, March 2024.

\bibitem{LPWA}
M.~Lauridsen \emph{et~al.}, ``{Coverage Comparison of GPRS, NB-IoT, LoRa, and SigFox in a 7800 km² Area},'' in \emph{2017 IEEE 85th Vehicular Technology Conference (VTC Spring)}, 2017, pp. 1--5.

\end{thebibliography}

\end{document}